\documentclass[aps,prd,nofootinbib,preprint,superscriptaddress,floatfix]{revtex4-1}
\usepackage{graphicx,color}
\usepackage{subfig}
\usepackage{amstext,amsmath,amssymb,amsfonts,amsthm}
\usepackage[format=plain,justification=centerlast]{caption}
\usepackage{mathrsfs}
\usepackage{hyperref}
\def\a{\alpha}

\def\e{\eta}
\def\lf{\left}
\def\rg{\right}
\def\g{\gamma}

\def\D{\Delta}

\def\fr{\frac}

\def\l{\label}
\def\G{\Gamma}
\def\no{\nonumber}
\def\be{\begin{equation}}
\def\ee{\end{equation}}
\def\ber{\begin{eqnarray}}
\def\eer{\end{eqnarray}}

\begin{document}

\author{Carlos Villalpando}
\email{cvillalpando@ucol.mx}
\affiliation{Facultad de Ciencias - CUICBAS, Universidad de Colima, Colima, C.P. 28045, M\'exico}
\affiliation{Fermi National Accelerator Laboratory, Batavia, Illinois 60510, USA}

\author{Sujoy K. Modak}
\email{smodak@ucol.mx}
\affiliation{Facultad de Ciencias - CUICBAS, Universidad de Colima, Colima, C.P. 28045, M\'exico}


\title{Indirect Probe of Quantum Gravity using Molecular Wave-packets}   

\begin{abstract}
The most obvious obstacle behind a direct test of Quantum Gravity (QG) is its energy scale ($10^{19}$ GeV), which remains well outside of any human made machine. The next best possible approach is to provide indirect tests on effective theories of QG which can be performed in a lower energy scale. This paper is aimed in this direction, and shows a promising path to test the existence of the fundamental minimal length scale of Nature by measuring the dispersion of free, large molecular wave-packets. The existence of the minimal length is believed to be the reason for a modified commutation relationship between the position and momentum operators and, in this paper, we show that such a modification of the commutator has a profound effect on the dispersion rate of free wave-packets, and precise measurement on the broadening times of large molecular wave-packets (such as $C_{60}$, $C_{176}$ and large organic molecules) provide a promising path for an indirect test of quantum gravity, in a laboratory setting.
\end{abstract}

\maketitle

The existence of quantum gravity (QG) theory is often associated with the existence of a fundamental minimal length (at Planck value $l_P = 10^{-35}$ m) in Nature which, however, is nothing more than a speculation. There are several proposals which are indicative of a {\it minimal length}, coming from string theory \cite{ST, st-garay}, black hole physics \cite{BH1, BH2}, Doubly Special Relativity \cite{DSR3}, Loop Quantum Gravity (LQG) \cite{LQG}, non-commutative geometries \cite{NC} and other, more general approaches \cite{gen}.  Further, one of the  consequences of this minimal length is believed to be a reason of replacing the  Heisenberg's Uncertainty Principle (HUP) by the so-called Generalized Uncertainty Principle (GUP) \cite{BH1, st-garay, gup-rev}. This assertion of a GUP with minimal length (or quantum gravity in a broader sense) has generated enormous interest in ``QG phenomenology'' \cite{bounds, Das-Vagenas, stb} and measuring GUP contributions has become a major task for the community \cite{exps}.

In this paper we provide a new avenue which may eventually allow us to test (indirectly) the existence of minimal length in Nature.  The case study for this is a  free particle wave-packet where the {\em{bare effect of minimal length}} will not mix with any other force fields. There exist few preliminary studies on the GUP effect on free particle wave-packet \cite{Nozari}, but in this paper we make crucial advancements  which lead to the construction of a new path to test the GUP theories by studying the expansion of large molecular wave-packets.

In quantum mechanics, the Heisenberg Uncertainty Principle (HUP) combined with the kinematical Ehrenfest's equations provide an important result that   the width of the free particle wave-packet  is always spreading (in space) over time. In fact, the free particle case comes as a special case of the dynamical equation \cite{Messiah}
 \be 
\fr{d^2 \xi}{dt^2} \approx \fr{4}{m} (\varepsilon - V_{cl}'' \xi),
\l{eqmas}
\ee
valid for a 1-dimensional wave packet $\Psi(q,t)$ with Hamiltonian $H = \fr{p^2}{2m} + V(q)$, and it holds for potentials depending up to quadratic power in $q$. In the above equation $\xi = (\D q)^2 = \lf< q^2 \rg> - \lf< q \rg>^2$ (the mean square deviation in position), $\varepsilon = \lf< H \rg> - E_{cl} = \fr{1}{2m} \e + \lf< V \rg> - V_{cl}$ (the difference between the expectation value of the Hamiltonian and its classical approximation) with $V_{cl} = V \big(\lf< q \rg>\big)$, and $\e = (\D p)^2 = \lf< p^2 \rg> - \lf< p \rg>^2$ (mean square deviation in momentum space). Note that in the classical approximation $\Psi(q,t)$ represents a particle with position, momentum and energy given by $q_{cl} = \lf< q \rg> ,~ p_{cl} = \lf< p \rg>~ \text{and}~E_{cl} = \fr{\lf< p \rg>^2}{2m} + V \big(\lf< q \rg> \big)$. For the classical approximation to hold, we require the extension $\D q$ of the wave packet to remain small as compared to the characteristic distances of the problem under consideration.

Upon solving \eqref{eqmas}, and knowing the deviations $\xi_{0}$, $\e_{0}$, and $\dot{\xi_{0}} \equiv d\xi_{0} /dt$ at $t = t_{0}$, we obtain $\xi(t)$, i.e., the spread of the wave function over time in configuration space; $\e (t)$ (spread in momentum space) can then be found using the fact that $\varepsilon$ is constant.

In the case of the free particle, $V = 0$, and thus we have $\e = 2m\varepsilon = \e_{0}$, that is, $\e = (\Delta p)^2$ remains constant, and we have from \eqref{eqmas} $d^2\xi/dt^2 = 2\e_{0}/m^2$ and thus
\be 
\xi (t) = \xi_{0} + \dot{\xi_{0}}t + \fr{\e_{0}}{m^2}t^2.
\l{fp}
\ee
This result is telling us that free wave-packets \textit{spread} indefinitely and, further, sets a limit for the time interval during which the classical particle analogy holds. If we have $\dot{\xi_{0}} = 0$ (e.g., the packet is {\em{minimally}} wide at $t_{0}$, so that, $\xi_{0}\e_{0} = \fr{1}{4}\hbar^2$) then \eqref{fp} is simplified as $\xi = \xi_{0} + \e_{0}t^2/m^2$ or, equivalently,
\ber
\D q (t) = \sqrt{\xi} (t) = \lf[ (\D q_{0})^2 + \lf( \fr{\D p_{0} t}{m} \rg)^2 \rg]^{1/2},
\l{fpex}
\eer
where $\D q_{0}$ and $\D p_{0}$ are the initial uncertainties in position and momentum space corresponding to the minimum-width wavepacket. It is a truly remarkable result and fundamental to our physical understanding of the quantum theory. This explains why we cannot see an electron as a localized object while classical objects seem to remain localized forever. In case of a free electron the quadratic term in \eqref{fpex} matches the initial width for $t=\fr{2\pi (\D q_0)^2}{c \lambda_e}$ (by using the minimum wave-packet uncertainty relation $\D q_0 \D p_0 = \hbar/2$ and the expression for the Compton wavelength $\lambda_e$ for an electron). Using $\lambda_e = 2.4 \times 10^{-12}~m$ and an initial width $\D q_0 \simeq 10^{-10}~m$ it is easy to check the second term in \eqref{fpex} equates the first term in $t\sim 10^{-16}~s$. That is, the wave-packet delocalizes (due to HUP) rather quickly. On the other hand, this would not happen with classical objects even if we wait for the entire age of the universe.

This paper intends to calculate the modification of this broadening rate when HUP is substituted by a GUP, understand the modification physically and propose experiments to measure the time difference between the two pictures. The form of a GUP commutator that we consider here is  \cite{Das-Vagenas}
\be
\big[ q_{i}, p_{j} \big] = i\hbar \Big\{ \delta_{ij} - \a \big(p\delta_{ij} + \fr{p_{i} p_{j}}{p} \big) + \a^2 (p^2 \delta_{ij} + 3p_{i}p_{j}) \Big\},
\l{GUP}
\ee
which is correct up to the second order in momentum. We call $\a = \a_{0}/m_{P} c$ the GUP parameter with $\a_{0}$ a dimensionless quantity, $m_{P}$ the Planck mass, and $c$ the speed of light. The quantity $m_P c = 6.52485$ kg.m/s is the Planck momentum. This form \eqref{GUP} of GUP is the most refined version in the sense that it forbids the problem of associating an infinite energy (or momentum) to a particle in the GUP picture, which after all should only be associated with a point-like particle, where the existence of the minimal length is not a pre-requisite \cite{DSR3}. However, the machinery that we build here is completely general and independent of the particular form of GUP commutator that one chooses to work with, but results will vary with different choices.

\begin{figure}[t!]
\centering
\includegraphics[width=0.91\textwidth]{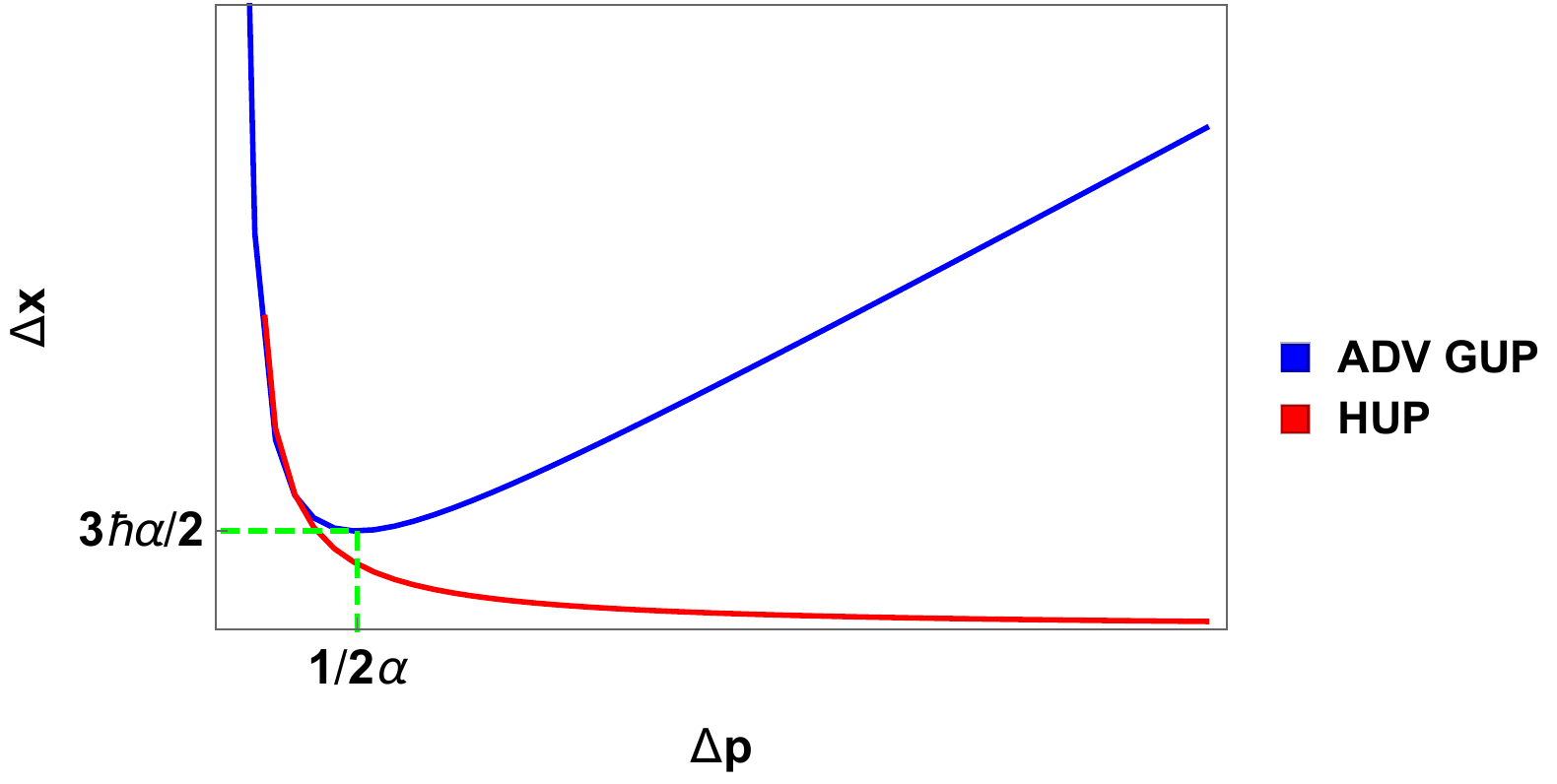}
\caption{Minimum uncertainty curve $\Delta x (\Delta p)$ for both the HUP (red line) and the GUP (blue line). The GUP implies a minimal length uncertainty $(\Delta q)_{min} = 3 \hbar \alpha / 2$ for the value $\Delta p = 1 / 2 \alpha$; this point is highlighted with the green dashed lines.}
\label{qmin-adv}
\end{figure}

We shall consider here a one-dimensional wave-packet, assuming that we are interested to measure the spread along a given axis, for which \eqref{GUP} becomes
\be 
\big[ q,p \big]_{GUP} = i \hbar(1 - 2\a p + 4\a^2 p^2 )
\l{gamma}
\ee
As a curious observation let us comment that we can get rigid bounds (both upper and lower) on the parameter $\a$ from the plot in Fig. \ref{qmin-adv}. Notice that the GUP predicts a minimal length $(\Delta q)_{min} = 3 \hbar \alpha /2$ for the value $\Delta p = (2 \alpha)^{-1}$, as shown in Fig. \ref{qmin-adv}.
This shows that, there is a connection between the size of the fundamental length with the value of the GUP parameter $\a$. Fixing one fixes the other automatically. For example, if by some experiment we can fix $\a = {\cal O} (1)$,  it would mean $(\Delta q)_{min} \simeq \ell_P$ and also if the minimal length cannot be lower than the Planck length one would associate a lower limit for $1\le\a$. Similarly, on the other hand, the particle accelerator experiments at LHC  we have already accessed a length-scale approximately around $10^{-18}$ m. This can again be related with $(\Delta q)_{min}$, now for obtaining an upper bound on $\a$. Noting $(\Delta q)_{min} = 3 \hbar \alpha / 2 \le 10^{-18} \implies \a \le 10^{16}$. Therefore, we obtain a hard bound on the GUP parameter given by $1\le \a \le 10^{16}$ just by knowing the fact that the fundamental length, if it exists, must be between $10^{-18}$ m and the Planck length $10^{-35}$ m.  We shall come back to this discussion once again at the end of our analysis.


With the definition \eqref{gamma} one can derive, following the identical steps as for \eqref{eqmas} (given in \cite{Messiah}), the defining equation for the expansion of free wave-packets in the GUP picture.\\ 
Using the identity
\begin{equation} \label{ident}
i\hbar \frac{d}{dt} \left< A \right> = \left< \big[ A, H \big] \right> + i\hbar \left< \frac{\partial A}{\partial t} \right>,
\end{equation}
(where $A$ is a Hermitian operator and $H$ is the system's Hamiltonian) along with the commutation relation \eqref{gamma} on $\xi = \lf< u \rg>$ (with $u = q^2 - \left< q \right>^2$) gives
\be \l{dgup}
\dot{\xi}_{GUP} = \fr{1}{m} \big( \lf< \g pq + q\g p  \rg> - 2q_{cl} \lf<  \g p \rg> \big)
\ee
where we have defined $\g \equiv 1 - 2\a p + 4\a^2 p^2$ as a simplifying notation. Repeating this procedure on \eqref{dgup}, and keeping terms up to $\a^2$ only, we obtain the {\em Generalized Master Equation}

\be
\begin{split}
\ddot{\xi}_{GUP} = & \fr{2}{m^2} \lf\{ \lf< \g p^2 \rg> - \lf< \g p \rg>^2 - 2\a \lf< \g p^3 \rg> + 4\a^2 \lf< \g p^4 \rg> \rg\} \\
& + \fr{1}{m} \bigg\{ 2q_{cl} \lf< \g V' \rg> - \lf< \g V'q + q\g V' \rg> \\
 & + 2\a \big[ \lf< \g \chi q \rg> + \lf< q\g \chi \rg> - 2q_{cl} \lf< \g \chi \rg>  \big] \\
 & - 4\a^2 \big[ \lf< \g \varpi q \rg> + \lf< q \g \varpi \rg> - 2q_{cl} \lf< \g \varpi \rg> \big] \bigg\} \hspace{0.3cm} .
 \end{split}
 \l{masterGUP}
 \ee

\vspace{0.15cm}
 
where $\chi = V'p + pV' \hspace{0.10cm} , \hspace{0.10cm} \varpi = V'p^2 + pV'p + p^2V'$ and $\g = 1 - 2\a p + 4\a^2 p^2$.\\
Restricting to the free-particle case ($\chi = \varpi = V = 0$) takes \eqref{masterGUP} to

\be
\ddot{\xi}_{free} = \fr{2}{m^2} \bigg\{ \e_{0} - 4\a \big[ \lf< p^3 \rg> - p_{cl}\lf< p^2 \rg> \big] + 4\a^2 \big[ 3\lf< p^4 \rg> - \lf< p^2 \rg>^2 - 2p_{cl}\lf< p^3 \rg> \big] \bigg\}, \l{d2fp}
\ee
where $\e_{0} = 2m\varepsilon$ (as found before). For a free particle it is easy to check that $\lf< p \rg> = p_{cl}$, $\lf< p^2 \rg>$ and all higher moments $\lf< p^n \rg>$ are constants in time. This makes \eqref{d2fp} easy to solve. The exact expression dictating the spread over time, assuming that the packet had a minimum width at $t = t_{0}$ (implying $\dot{\xi}(t_{0}) = 0$), is
\ber
&&\D q_{free}(t) = \sqrt{\xi} (t) \nonumber\\
&& = \sqrt[]{{\D q_{0}}^2 + \fr{1}{m^2} \big( {\D p_{0}}^2 - 4\a C_{1} + 4\a^2 C_{2} \big) t^2 } ,
\l{fpgup}
\eer
where the terms coupled with the GUP parameter are $C_{1} = \lf< p^3 \rg> - p_{cl}\lf< p^2 \rg>$ and  $C_{2} = 3\lf< p^4 \rg> - \lf< p^2 \rg>^2 - 2p_{cl}\lf< p^3 \rg>$.

To understand the new element brought in by the GUP we need an interpretation of the coefficients $C_1$ and $C_2$. First thing to note is that they involve \emph{higher-order moments} and thus introduce a novel statistical interpretation concerning the shape of the probability distribution in momentum space. It is useful to introduce \emph{Pearson's skewness coefficient} for the third-order moment, as
\be 
\G_{1} = \fr{\lf< \lf( p - \lf< p \rg> \rg)^3 \rg>}{\sigma^3} = \fr{1}{\e^{3/2}} \lf< \lf( p - \lf< p \rg> \rg)^3 \rg>,
\l{skewness}
\ee
whereas, the fourth order moment is given by the \emph{kurtosis coefficient}
\be
\G_{2} = \fr{\lf< \lf( p - \lf< p \rg> \rg)^4 \rg>}{\sigma^4} = \fr{1}{\e^2} \lf< \lf( p - \lf< p \rg> \rg)^4 \rg>.
 \l{kurtosis}
\ee
The term $\sigma \equiv \sqrt[]{\lf< p^2 \rg> - \lf< p \rg>^2} = \e^{1/2}$ is the \emph{standard deviation} in momentum distribution, which also appears without the GUP modifications. Usually $\G_1$ and $\G_2$ measure the departure of probability distributions from the \emph{normal distribution}.  While $\G_1$ measures the asymmetry about the mean $\lf< p \rg>$, $\G_2$ measures its tailed-ness. Pearson's skewness can take positive or negative values, but kurtosis is a positive definite quantity. A normal (or true Gaussian) distribution is characterised by $\G_1 = 0$ and $\G_2 = 3$. 

In a slightly expanded form these coefficients are
\begin{equation*}
\Gamma_1 = \frac{1}{\eta^{3/2}} \left( \left< p^3 \right> + 2\left< p \right>^3 - 3\left< p \right>\left< p^2 \right> \right),
\end{equation*}
and
\begin{equation*}
\Gamma_2 = \frac{1}{\eta^2} \left( \left< p^4 \right> - 4\left< p \right>\left< p^3 \right> + 6\left< p^2 \right>\left< p \right>^2 - 3\left< p \right>^4 \right).
\end{equation*}

Using these we can write
\begin{equation} \label{C1}
C_1 = \eta \left( 2 p_{cl} + \Gamma_1 \eta^{1/2} \right) ,
\end{equation}
and
\begin{equation} \label{C2}
C_2 =  \left( 3\Gamma_2 - 1 \right) \eta^2 + 10 p_{cl} \eta \left( \Gamma_1 \eta^{1/2} + p_{cl} \right)
\end{equation}
which, apart from the standard deviation (i.e. $\e$) also include skewness, kurtosis of the momentum distribution. It is an important result - it says the broadening rate for free wavepackets depends on the shape of the probability distribution in momentum space. Different wavepackets with same standard deviation but different skewness or kurtosis have different broadening rates with the GUP modification. On the other hand a HUP based calculation is completely blind to this difference. 

For  an account of the broadening rate with GUP and a comparison with HUP we need to express $\e_0$ (which is constant over time) as a function of the wave-packet's initial \textit{size} $\xi_0 = \lf( \D q_0 \rg)^2$. We can find this using the minimum uncertainty relation
\ber 
\lf( \D q_0 \D p_0 \rg)_{GUP} = \fr{\hbar}{2} \lf[ 1 + \lf( \fr{\a}{\sqrt{\lf< p^2 \rg>}} + 4\a^2 \rg) \D p_0^2 \rg.\nonumber\\
\lf. + 4\a^2 p_{cl}^2 - 2\a \sqrt{\lf< p^2\rg>} \rg] 
\l{mun}
\eer
Using this, and $\lf< p^2 \rg> = \e_0 + p_{cl}^2$, we find the following expression
\be 
\fr{2}{\hbar} \lf( \D q_0 \sqrt{\e_0} \rg) - \lf[ 1 + 4\a^2 \lf(\e_0 + p_{cl}^2 \rg) \rg] + \a \lf[ \fr{\e_0 + 2p_{cl}^2}{\sqrt{\e_0 + p_{cl}^2}} \rg] = 0
\l{feta}
\ee
Upon solving \eqref{feta} we obtain the expression $\e_0 = \e_0 \lf( \D q_0, \a, p_{cl} \rg)$. As we mentioned, it is easy to see $\e_0$ is constant in time since both $\D q_0$ is the initial spread and $p_{cl}$ is also a constant in time.

The results we obtained so far allow for an experimental verification of the minimal length effect by measuring the timescale in which the wave-packet (associated with a particle or a system of particles) doubles its initial width. In fact, one may choose any final size that is permissible by experimental set up, but we show our calculation considering the wave-packet doubles its size. 

This ``doubling time'' with a HUP based calculation was already discussed for the case of electrons below eq. \eqref{fpex}. Likewise, we can express this in the GUP framework  using \eqref{fpgup}
\be
t_{\text{double}} = \fr{\sqrt{3} m \D q_0}{\sqrt{\D p_0^2 - 4\a C_1 + 4\a^2 C_2}},
\l{doub}
\ee
where the minimum-uncertainty wave-packet has to satisfy \eqref{mun}. The definitions of $C_1$ and $C_2$ then give an estimate for $t_{\text{double}}$. Obviously, different distributions with different $\G_1$ and $\G_2$ will give different values of this duration of time even if these distributions have the same variance. This is a striking departure from HUP results. Even if we consider a distribution (or Gaussian wavepacket) with  $\G_1 = 0$ and $\G_2 = 3$ we can obtain measurable differences in this durations with or without the GUP. For a normal distribution
\be
t_{\text{double}}= \fr{\sqrt{3} m \D q_0}{\sqrt{\D p_0^2 + 8 \e \big( \a^2 \lf( 4\e + 5 p_{cl}^2 \rg) - \a p_{cl} \big)} }
\l{doub-n}
\ee
Using above expressions along with \eqref{feta} we can now replace $\e = \D p_0 ^2$ in terms of $\D q_0$ and the other parameters. This brings us to a position of making a numerical analysis of the results.

The example of free electrons discussed before is a good place to start with. Considering the initial width of the wave-packet as $10^{-10}~m$, we can use \eqref{doub-n} to estimate that, for a range of GUP parameter $1 \le \a \le 10^{21}$, the doubling time  essentially remains of the same order of magnitude, with or without the GUP (given by $t_{\text{double}} \simeq 10^{-16}~s$). In fact, for relatively smaller values of the GUP parameter such as $\a \leq 10^{10}$ the difference between GUP and HUP doubling times is negligibly small ($10^{-30} ~s$). Whereas, this difference becomes  ${\cal O}(10^{-19}) ~s$ for $\a \sim 10^{21}$. These differences are too small to detect even with state-of-the-art atomic clock technology. Furthermore, if the stringent bound coming from the precision observations of the Lamb shift needs to be respected (implying $\a \le 10^{10}$ \cite{stb}), these differences in doubling time for free electrons are unattractive.

To make the case more attractive, so to speak, and providing a possible experimental verification, we have to consider sizes of initial wave-packets larger than that of free electrons. One straightforward way of doing that is to consider atomic and molecular wave-packets. There are many examples that one can borrow from the literature and we choose some of the most studied ones - such as the ``buckyballs''  and Large Organic Molecular (LOM) wave-packets.

Buckyballs are basically closely packed Carbon atoms behaving as a single quantum mechanical wave-packet \cite{c60}. For example, a $C_{60}$ {buckyball} molecule has  mass $1.19668 \times 10^{-24}$ kg (720 amu) and the initial width of the wave-packet $\D q_0$ can be considered as equal to its van der Waals diameter (7 \AA) \cite{buckysize}. The HUP based calculation predicts a doubling time $t_{\text{double}} = 1.92719 \times 10^{-8} s$. The GUP based calculation has a free parameter $\a$ and the doubling time varies considerably for various values of this parameter. For $\a \sim {\cal O} (1)$ the difference from HUP is practically unmeasurable. However, for larger values like $\a = 10^{10}$,  this difference $\D t_{\text{double}} = t_{\text{double}} (GUP) - t_{\text{double}} (HUP)$ is given by
\be \no
\D t_{\text{double}} (C_{60}, \a = 10^{10}) = 1.15631 \times 10^{-14} s.
\ee
This difference in time, however, stays within the bound of precise atomic clocks which can measure times of the order of femto-seconds.

Let us consider now a $C_{176}$ molecule for which  $m = 3.50706 \times 10^{-24}$ kg (2112 amu) and $\D q_0 = 1.2$ nm \cite{buckysize}. The HUP predicted doubling time is $t_{\text{double}} (C_{176}, HUP) = 1.6598 \times 10^{-7} s$ and, the difference between the HUP and GUP predictions for $\a = 10^{10}$ is
\be \no
\D t_{\text{double}} (C_{176}, \a = 10^{10})  = 9.9588 \times 10^{-14} s.
\ee
This is an improvement by almost one order of magnitude.

This difference is magnified to a relatively higher proportion with large organic molecular wave-packets \cite{LOM}. The {\em Tetraphenylporphyrin} or TPPF152 molecule ($C_{168}H_{94}F_{152}O_{8}N_{4}S_{4}$) is one such molecule consisting of 430 atoms. It has a  mass of $5,310$ amu ($8.81746 \times 10^{-24}$ kg) and the initial width of the wave-packet could be considered of 60 \AA. For $\a =10^{10}$ the time difference between the HUP and GUP results is
\be \no
\D t_{\text{double}} (\text{TPPF152}, \a = 10^{10})  = 6.25961 \times 10^{-12} s.
\ee
This is a remarkable improvement of the result of  $C_{176}$ (by a factor of 63) and  $C_{60}$ molecule (by a factor of 500!). 

\begin{figure}[t!]
\includegraphics[width=\textwidth]{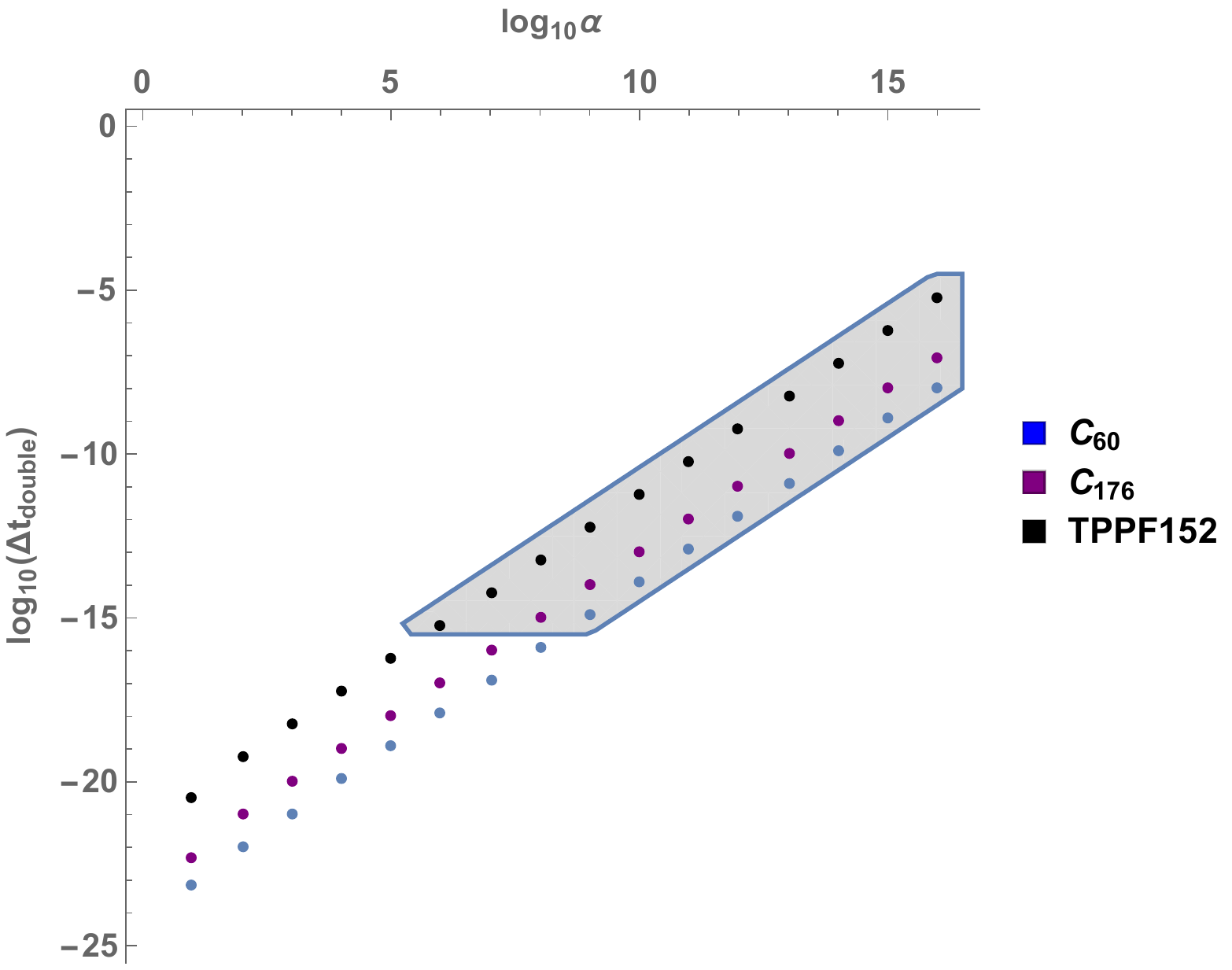}
\caption{Logarithmic plot between the ``doubling time'' difference (between the GUP and HUP) and the GUP parameter $\a$ for (a) the $C_{60}$ molecule (lower), (b) the $C_{176}$ molecule (middle), and (c) TPPF152 (upper). We have shaded the region of the parameter space which can  be probed by these molecular wave-packets with the maximum precision of the atomic clock set as $ \sim 10^{-15}~s$, quite conservatively.}
\l{ll}
\end{figure}

A summary of results for the doubling time difference for physically interesting values of the GUP parameter ($1\le \a \le 10^{16}$) is plotted in Fig. \ref{ll}. Both of the axes are scaled logarithmically. The difference between the doubling time for the lowest allowable value $\a \sim 1$ is beyond our computer precision - hence this is not included in Fig. \ref{ll}. For the rest $10 \le \a \le 10^{16}$ the $Y$ axis gives the difference of two doubling times which is dependent on the corresponding values of the GUP parameter $\a$ plotted along the $X$-axis. Expectedly, for the larger values of $\a$ the difference in time is quite large (of the order of microseconds for $\a \sim 10^{16}$ for TPPF152 molecule). For the smaller values of $\a$ the doubling time also become smaller and harder to detect. If we use atomic clocks with the available time resolution of femto-seconds ($10^{-15}$ s) we can access the shaded region of the parameter space, which, once again, is best for TPPF152 organic molecule - we can scan down to $\a \sim 10^6$. This is a four orders of magnitude improvement on the existing upper bound on $\alpha \le 10^{10}$ as found from the Lamb shift \cite{stb}. In summary, for the case of not detecting these differences with femto-second time resolution, the new upper bounds  with $C_{60}$, $C_{176}$ and  TPPF152 molecules will be $\a  \le10^{9}$,  $\a \le10^{8}$, and $\a \le10^{6}$, respectively. On the other hand a successful detection will fix this parameter $\a$ for \eqref{gamma}.
 
Now coming back to the discussion of the value of $\alpha$ and the size of the fundamental length scale we see that an observation of $\a \sim 10^6$ would fix the fundamental minimal length $(\Delta q)_{min}\sim10^{-29}$ m as follows from the discussion around Fig. \eqref{qmin-adv}. This is still 11-12 orders of magnitude lower than the length scale probed so far by high energy  accelerator experiments at the LHC and therefore cannot be overruled by any means. However, non-detection of  the said doubling time difference in the above molecular wave-packet experiments would suggest that the fundamental minimal length scale is even below $10^{-29}$ m and we need to improve our set up to further probe the region of $1\le \a \le 10^{6}$.
 
Indeed, it is fascinating to note that large molecular wave-packets which are usually used to probe the classical-quantum interface (by studying the interference pattern in double slit experiments \cite{c60, LOM}) may also be invaluable to obtain an indication on the minimal length scale of Nature. In this paper, we have discussed that GUP  proposal such as \eqref{GUP} not only brings a rich distributional modification on the expansion rate of free wave-packets but also modifies the ``doubling time'', that is, the time in which a free, minimal width wave-packet doubles its size. This difference in broadening time is bigger for massive molecular wave-packets in comparison with the wave-packets representing smallest objects like electrons. Further improvements to our results to scan {\it all} parameter space for $\a$, summarized in Fig. \ref{ll}, are quite possible and ways to do so are two folded - such as - (i) to consider larger and heavier molecular wave-packets (more than the LOM in \cite{LOM}), and (ii) to come up with even more precise atomic clocks (with an ability to measure a time difference beyond a femto-second). This is a new avenue that has not been proposed before and  we expect, perhaps, experimentalists will be interested in taking this path.  We find this aspect particularly interesting which may lead us to indirect evidence for the existence or non-existence of a fundamental minimal length or, to a broader sense, a theory of  quantum gravity.

{\bf Acknowledgements:} CV thanks Fermi National Accelerator Laboratory for allowing him the office space and research facilities during his stay when a major part of this research work was completed. Research of SKM is supported by the CONACyT Project No. CB17-18/A1-S-33440 (Mexico).


\end{document}